\documentclass[apj]{emulateapj}   
\usepackage{graphicx}
\usepackage{multirow}

\usepackage{amsmath}
\usepackage{natbib}
\usepackage{hyperref}
\usepackage[usenames,dvipsnames]{color}  

\def\msun{{\rm M}_{\odot}}
\def\rsun{{\rm R}_{\odot}}
\def\Mpc{{\rm Mpc}}
\def\Gpc{{\rm Gpc}}
\def\myr{{\rm Myr}}

\def\mpy{{\rm M}_{\odot} {\rm ~yr}^{-1}}

\newcommand{\kms}{\ensuremath{\,\rm{km}\,\rm{s}^{-1}}}
\newcommand{\perMyr}{\ensuremath{\,\rm{Myr}^{-1}}}
\newcommand{\Msun}{\ensuremath{\,M_\odot}}

\newcommand{\Rsun}{\ensuremath{\,R_\odot}}
\newcommand{\kpc}{\ensuremath{\,\rm{kpc}}}

\begin{document}

\title{Merger rates of double neutron stars and stellar origin black
holes: The Impact of Initial Conditions on Binary Evolution Predictions}

\author{S. E. de Mink\altaffilmark{1} and K. Belczynski\altaffilmark{2,3}} 

 \affil{
     $^{1}$  Anton Pannenkoek Astronomical Institute, University of
Amsterdam, 1090 GE Amsterdam, The Netherlands (S.E.deMink@uva.nl)\\
     $^{2}$ Astronomical Observatory, Warsaw University, Al.
            Ujazdowskie 4, 00-478 Warsaw, Poland
            (kbelczyn@astrouw.edu.pl)\\
     $^{3}$ Warsaw Virgo Group\\
         }

\begin{abstract}

The initial mass function (IMF), binary fraction and distributions of binary parameters (mass ratios, separations and eccentricities) are indispensable input for simulations of stellar 
populations. It is often claimed that these are poorly constrained significantly affecting evolutionary predictions. 
Recently, dedicated observing campaigns provided new constraints on the initial conditions for massive stars. Findings include a larger close binary fraction and a stronger preference for very tight systems. We investigate the impact on the predicted merger rates of neutron stars and black holes. 
Despite the changes with previous assumptions, we only find an increase of less than a factor 2 (insignificant compared with evolutionary uncertainties of typically a factor $10-100$).  
We further show that the uncertainties in the new initial binary properties 
do not significantly affect (within a factor of $2$) our predictions of double 
compact object merger rates.  An exception is the uncertainty in IMF 
(variations by a factor of $6$ up and down). No significant changes in the 
distributions of final component masses, mass ratios, chirp masses and delay 
times are found.
We conclude that the predictions are, for practical purposes, robust against 
uncertainties in the initial conditions concerning binary parameters with  
exception of the IMF. This eliminates an important layer of the many uncertain 
assumptions affecting the predictions of merger detection rates with the 
gravitational wave detectors  aLIGO/aVirgo.

\end{abstract}

%
\section{Introduction}
These are very exciting times for gravitational wave astrophysics.  
The direct detection of the gravitational wave signal
of the merger of two compact objects, neutron stars (NS) or black holes
(BH) is anticipated before the end of this decade. 

Gravitational waves are a natural consequence of the theory of General
Relativity \citep{Einstein1918}. They are perturbations of the spacetime 
metric propagating at the speed of light, which are generated, for example, 
during the inspiral of two compact objects. 
Indirect evidence for the existence of gravitational waves was provided
by the orbital decay of the Hulse-Taylor pulsar \citep{Hulse+1975,
Taylor+1989}  and later with stronger constraints by the double pulsar 
\citep{Burgay+2003,Lyne+2004}. Direct detection of the gravitational wave signal of the
inspiral of NS-NS, BH-NS or BH-BH binaries and the subsequent merger and
ring down is expected to happen in the next few years, now the advanced
ground-based gravitational wave detectors aLIGO and Virgo are coming
online \citep{Abbott+2009, Caron+1997}.  
 
The initial LIGO/Virgo observations were concluded in 2010 without
detection, but they provided upper limits on the merger rates
\citep{Abadie+2012}. The advanced version of the detectors
will be approximately 10 times more sensitive than the initial versions,
expanding the detection volume and thus the chance of detection by a
factor of about a thousand \citep{Aasi+2013}.

The first science run of advanced LIGO is scheduled for late 2015 \citep{Aasi+2013a}.  A
detection during the first science run is not considered to be likely, but one or
more detections are anticipated in the next few years as the sensitivity
increases to a range of 200 Mpc for double neutron stars. This
translates to $\sim 0.2$--$200$ expected detections double neutron star mergers
per year \citep{Aasi+2013a}.  Even without
detections, the new upper limits will become astrophysical interesting
as they start to rule out the models that predict the highest merger
rates \citep{Mandel+2010,Belczynski+2012,Stevenson+2015}.

Obtaining reliable predictions of the merger rates of relativistic
compact objects has been a very large challenge, as reviewed by \citep{Abadie+2010}. 
The rates quoted above
are derived semi-empirically using the observed binary pulsars in our
Galaxy \citep{Phinney1991,Kalogera+2004}. The large uncertainties result
from the small size of the sample of observed binary pulsars and the pulsar luminosity
distribution. The implicit assumption is made that the observed sample
is representative for the Galactic population. Further constraints come from 
short gamma-ray bursts, but the rate estimates depend on the uncertain luminosity function  and opening angle of the jets \citep[e.g.][]{Fong+2012}. Type Ibc supernovae have served so 
far as ultimate upper limits \citep{Kim+2010}.  The merger rates for BH-NS and BH-BH
systems are even more problematic as we lack direct observational
evidence of their existence. However, some immediate progenitors for 
BH-BH \citep[e.g.][]{Bulik+2011} and BH-NS systems \citep[e.g.][]{Grudzinska+2015} 
have been proposed. 

Even though theoretical predictions suffer from even larger uncertainties, they have 
been crucial to estimate the merger rates involving black holes. They further provide 
the expected distribution of properties for all merger types including for example 
the delay times, component and chirp masses. More importantly, the theoretical 
predictions are providing the tool for future comparison with the detections, crucial 
to identify what gravitational wave signals teach us about the astrophysics of the 
progenitor systems \citep[e.g.][]{Stevenson+2015}. 
Several groups have presented similar estimates and studies in the past
decade  \citep[e.g.][]{Lipunov+1997,Bethe+1998,De-Donder+1998,Bloom+1999,
Grishchuk+2001,Nelemans+2001,Voss+2003, Dewi+2003,Nutzman+2004,De-Donder+2004,
Pfahl+2005,Postnov+2006,Yungelson+2006,OShaughnessy+2008,OShaughnessy+2010,Dominik+2012, 
Dominik+2013, Mennekens+2014,Dominik+2015}.

For all predictions we can distinguish several sources of
uncertainty: (I) the adopted \emph {initial conditions} in the
simulations, (II) the uncertainties in the \emph{physics of
the stellar evolution and binary interaction},  (III) the uncertainties associated
with the \emph{normalization} of the merger rates, which include the
mapping to the appropriate star formation history of the detection
volumes and (IV) modeling of the \emph{detectability} of gravitational waves from the predicted merger events.

Recent observations have provided new constraints on the initial
conditions and primordial binary properties \citep{Kobulnicky+2007, Kiminki+2012,
Chini+2012, Sana+2012, Sana+2013, Sana+2014, Kobulnicky+2014, Moe+2015, Moe+2015a, Dunstall+2015}, for a review see \citep{Duchene+2013}. 
The studies show that the initial conditions for young massive stars differ substantially from the initial conditions that
have typically been adopted to simulate compact object mergers coming
from binary evolutionary channels.

Among the most striking findings are (i) the large fraction massive stars
that have a companion close enough to interact by exchanging mass before
they die and (ii) the preference for very tight binaries with orbital
periods of a few days, implying that a large fraction of massive stars will
interact even before leaving the main sequence. Further findings include
(iii) the confirmation of a flat distribution of mass ratios, ruling out a
distribution that is strongly peaked to equal masses and (iv) a distribution
of eccentricities that favors circular systems in stark contrast with the
typically adopted thermal eccentricity distribution which favors eccentric
systems.

The first two conclusions depend on adopted model of stellar evolution and in particular they require at least modest radial expansion for the majority of massive  stars. The amount by which stars expand is considerably uncertain for high mass stars. Note that some very rapidly rotating massive stars may actually decrease in size as they evolve \citep{Yoon+2005}, possibly even preventing interaction with  companion by mass transfer \citep{de-Mink+2009a} in very tight systems. 
For example, let us consider the fraction of massive binary systems in which mass transfer starts already during the main sequence evolution of the primary star. This depends on the maximum radius that a star reaches on the main sequence which depends on a number of rather uncertain processes such as convection and overshooting, \citep[e.g. calibrations by][]{Pols+1997, Ribas+2000, Brott+2011}, further mixing processes such as those induced by rotation \citep[e.g.][]{Yoon+2005, Brott+2011a, Ekstrom+2012, Szecsi+2015}, mass loss by stellar winds and eruptions \citep[review by][]{Smith2014} and the possible density inversions in the outer layers of massive stars which can result in inflated envelopes\citep[e.g.][ and Jiang et al. 2015, subm.]{Yusof+2013, Kohler+2015}. 
Varying the maximum radius of main sequence stars by $30\%$ up and down and allowing for uncertainties in the initial distributions, we find a large variation for this fraction, $20\%-50\%$\footnote{This estimate was done with the {\tt StarTrack} population synthesis code described in Section 3 considering systems with primary masses in the range $15-150\msun$. The initial distributions of period and eccentricity were altered within limits listed in Section 2.1 assuming a binary fraction of $100\%$.}. 

 A further caveat to keep in mind is that the observations are limited to regions nearby such as our own Galaxy and the Magellanic Clouds which only cover metallicities down to about one fifth of the solar value.  Also the regime of the highest mass stars is not well probed.  \citet{Sana+2012} observations cover stars in mass range $15 - 60 \msun$, but most of these are towards the lower end of this mass range.  For low metallicities and higher star masses we have no direct constraints and the uncertainties in the evolutionary models start to play a larger role. Despite several attempts, no trends with metallicity or environment are found so far \citep[e.g.][]{Bastian+2010, Moe+2013}.  We extend \citet{Sana+2012} distributions in our study to the entire range that can produce double compact objects ($M_{\rm zams}=5 -150 \msun$) and to a broad range of metallicities ($Z=0.002 - 0.02$) as discussed in Section 2.

In this study we focus on the binary formation channels  \citep[in contrast to
the dynamical formation channels that may occur in dense star clusters, 
e.g.][]{Ivanova+2008,Banerjee+2010,Aarseth2012,Clausen+2013,Samsing+2014,Ramirez-Ruiz+2015}.
We investigate two questions: (i) What are the implications of the new
initial conditions? and (ii) How robust are the predictions against the
allowed variations in the new initial conditions resulting from the
observational uncertainties.  

For this purpose we perform a comparative population synthesis study
where we use the recent work by \citet{Dominik+2012} as a reference.  We
simulate the evolution of massive binary systems following the
evolutionary channels for the formation of double compact objects.  In
Sect.~\ref{sec:initdist} we describe the new and old initial conditions.
 In Sect.~\ref{sec:method} we give a brief description of our
computational method, the physical assumptions and computation of the
merger rates. In Sect.~\ref{sec:dist} we compare the old and new initial
distributions for the entire simulated populations and for the
progenitors of double compact object mergers.  In Sect.~\ref{sec:rates}
we discuss the impact on the resulting merger rates. Finally, in
Sect.~\ref{sec:concl} we present our discussion and conclusions.

\section{ Initial Distributions}\label{sec:initdist} 
\subsection{New standard model (N) and its variations\label{sec:initnew}}

Recent dedicated observing campaigns have provided new constraints on
the binary properties of young massive stars.  Here, we investigate the
impact of the distributions obtained by the work of \cite{Sana+2012}. 
This study is based on an intense spectroscopic monitoring spanning a
decade surveying all O-type stars in six nearby  ($\lesssim 3\kpc$)
very young (about 2\,Myr old) open star clusters and associations.  
 Even though the sample may seem of modest size it exceeds previous studies in 
completeness in terms of the fraction of systems for which orbital
solutions have been obtained.  It provided an average of 20 radial
velocity measurements for all 71 systems in the open clusters/associations that 
contain at least one O-star. 
This dataset includes orbital solutions for several long period systems (between 
100 and 1000 days) which are very challenging as they require a long term 
observing campaign.  This sample allowed a robust derivation of the underlying 
distribution of binary parameters after correction for incompleteness and biases. 

The very young ages, relatively low densities and velocity dispersion of the 
stars, imply that the effects of stellar
evolution and dynamical interactions are minimal. This makes this sample
the most suitable to provide constraints on the primordial binary
properties and thus the initial conditions for our simulations.  

The primary stars in this sample have spectral types ranging from O9.7 to
O3, which correspond approximately to a mass range from 15 to 60
\Msun.  This is appropriate for the progenitor systems of double compact
object mergers that involve at least one black hole.  Lower mass
binaries may dominate the formation of double neutron stars.  

The most suitable study for this mass range is provided by  \citet{Dunstall+2015}  for the early B-type stars in the 30 Dor region and the Cygnus OB2 sample analyzed by 
\citet{Kobulnicky+2014}, which contains stars down to spectral types of  B2.5V, 
which approximately corresponds to 8\Msun.  \cite{Wright+2015} infer an age spread 
for this region of 1-7 Myr. Statistically the findings by \citet{Kobulnicky+2014} and \citet{Dunstall+2015} are consistent with the distributions by \citet{Sana+2012}, although the results 
by \citet{Kobulnicky+2014} and \citet{Dunstall+2015} favor a flatter period distribution and a lower close binary fraction, similar to the old initial distributions that we use 
as reference (see Sec.~2.2). Whether this is a sign of a trend with decreasing 
primary mass, or whether the distribution of ages play a role is not clear. 

Sourcing from our conclusions, we find that the
changes in the period distribution have a rather small impact on
double compact merger rates. Thus we apply the \citet{Sana+2012} 
distributions as our new initial conditions for both O and early B stars 
for consistency. 
  
%
\paragraph{Orbital periods}  For the distribution of orbital periods,
$p$, we adopt the distribution \citet{Sana+2012} which significantly
favors short period systems.  Such a preference had been observed in
previous surveys uncorrected for biases \citep[e.g.][]{Mason+2009}, but
was generally interpreted as being the result of selection effects.
\citet{Sana+2012} demonstrated that this preference remains even after
carefully correcting for observational biases.   We adopt
\begin{equation}
 f_p (\log p)  \propto ( \log p )^\pi, \quad \quad \text{for $ \log p\in
[0.15, 5.5]$}
\end{equation}
 where $p$ is given in days. For our standard simulation we adopt 
$\pi=-0.5$. We change the slope of orbital period distribution from 
$\pi=-0.75$ in model N-p1 to $\pi=-0.35$ in model N-p2.

Note that the spectroscopic observations can only reliable  probe systems 
with $\log p \lesssim 3.5$.  However, wider systems can still produce double 
compact object mergers as we will show in the following section.  For wider 
systems we adopt the simplest assumption we can take and extrapolate the 
distribution, since we have no reasons to believe that the binary fraction 
suddenly drops beyond $\log p = 3.5$. 

This assumption is consistent with the recent findings by interferometric 
studies of nearby Galactic massive stars. \citet{Sana+2014} provided a large 
systematic survey probing companions of O stars at angular separations between 1 and 100 mili-arcseconds. For unevolved massive stars (O stars with luminosity class V)
the detected companion fraction reaches 100\% at 30
mili-arcseconds. The physical separations this corresponds to will
remain uncertain until more accurate distance
measurements become available. Roughly it corresponds to physical separations of
60-6000 AU. In our preferred units for the orbital separation $a$ this
corresponds to $\log a (\Rsun) =\,$4.1-6.1. Considering systems with typical masses we find that our extrapolation of the orbital period distribution and binary fraction are consistent with these observations. 

\paragraph{Mass ratios} For the distribution of mass ratios, which we
define as the mass of the initially less massive star over the mass of
the more massive star, i.e. $q \equiv M_2/M_1$, we use 
\begin{equation}
f_q (q)  \propto q^\kappa, \quad \quad \text{for $q \in [0.1, 1]$}
\label{eq:iqfNew}
\end{equation}
where $\kappa = -0.1 \pm 0.6$ according \citet{Sana+2012}.  We adopt
$\kappa = 0$ such that the distribution becomes uniform distribution,
which has also been found in several recent studies such as 
\citet{Kobulnicky+2007} and \citet{Kobulnicky+2014}.  To consider the
uncertainties we consider lower and upper limits of $\kappa=0.5$ in
model N-q1 to $\kappa=-0.7$ in model N-q2. 

We note that the most recent observations rule out the presence of a so-called 
twin population of equal mass systems \citep{Pinsonneault+2006}. The idea of
the possible existence of such a population gained interest as it favors
the formation of double compact object mergers, in particular through the
so-called double core formation channel proposed by \citet{Dewi+2006}.
However, the claimed evidence of such a population has been demonstrated
to be the result of observational biases towards equal mass systems that
were not accounted for \citep{Lucy2006, Sana+2013, Cantrell+2014}.   

\paragraph{Eccentricities}

For the eccentricity distribution in the very young open clusters \citet{Sana+2012} 
finds
\begin{equation}
f_e (e) \propto  e^\eta, \quad \quad \text{for $e \in [0.0, 0.9]$}
\end{equation}
with $\eta=-0.42\pm0.17$.  The very short period systems ($p \lesssim 4$~days) show 
a larger degree of circularization as expected from the short time scale for tidal 
circularization \citep{Zahn1975}. Unfortunately the data sample is not large enough 
to provide a reliable separation dependent eccentricity distribution.  Instead we 
adopt this distribution as initial distribution for our 
simulations, independently of the period.  We explicitly follow the effects of tides in our simulations, which quickly circularizes the shortest period systems, consistent with the observations.  As we will show later this assumption is justifiable as the variations in the 
eccentricity result in only minimal changes in the rates. 
In our standard simulation N we adopt ${\eta =
-0.42}$. We consider uncertainties by changing the the slope to
$\eta=-0.59$ in model N-e1 to $\eta=-0.25$ in model N-e2.

\paragraph{Binary fraction} The spectroscopic survey by
\citet{Sana+2012} yields a binary fraction of $f_{\rm SB} = 0.7\pm0.1$
after correcting for biases. The binary fraction  here is defined as 
\begin{equation}
f_{\rm bin} \equiv \frac{N_{\rm bin}}{N_ {\rm bin} + N_{\rm single}  }
\end{equation}
where $N_{\rm bin}$ and $N_{\rm single}$ are the number of binary
systems and the number of single stars respectively. 

This fraction refers only to systems that have orbital parameters within the
considered boundaries , i.e. $q \in [0.1, 1]$,  $\log p {\rm {(days)}}
\in [0.15, 3.5]$ and $e \in [0.0, 0.9]$. The remainder of the sample ($\sim 30\%$) 
consists of stars of unknown nature, including wider binaries, binaries with more 
extreme mass ratios and possibly genuine single stars. 
Motivated by the interferometric survey \citep{Sana+2014} we designate the remainder 
as wide binaries. We extrapolate the original \cite{Sana+2012} period distribution that
extends to $\log p  = 3.5$ to $\log p = 5.5$. Such extrapolation results in a binary 
fraction, including wide systems, of $f_{\rm bin} = 1$.   We also consider a reduced 
binary fraction of $f_{\rm bin}=0.85$ in model N-f1 and  $f_{\rm bin}=0.7$  in model N-f2.

\paragraph{Masses} The observed distribution of primary masses in the
sample of  \citet{Sana+2012} is consistent with a standard 
\citet{Kroupa2002} mass function.  Although we only simulate the
evolution of massive binaries, for normalization we adopt the three
component power law for the primary mass or single star $m_1$, given in
solar units \Msun
\begin{equation}
f_{m_1}(m_1) \propto \begin{cases}
m_1^{-1.3}, &\text{for   $ m_1 \in [0.08, 0.5]$}\\
m_1^{-2.2}, &\text{for $ m_1 \in [0.5, 1.0]$}\\
m_1^{-\alpha}, &\text{for $m_1 \in  [1,150]$}
\end{cases}
\label{eq:imf}
\end{equation}
We adopt $\alpha = 2.7$ in our standard model, consistent with the field
star population \citep{Kroupa+1993, Kroupa+2003}. To allow for
uncertainties we consider $\alpha=3.2$ in model N-m1 and $\alpha=2.2$ in
model N-m2.

%
%
\subsection{Old standard model (O) and its variations \label{sec:initold}}

We adopt the \citet{Dominik+2012} reference model as the old standard model O.
In their work systems are drawn from an initial distribution of separations 
instead of orbital periods. This distribution is assumed to be flat in the 
log \citep{Abt1983,Opik1924}
\begin{equation}
 f_a (\log a)  \propto {\rm constant},  \quad \quad \text{for $ \log
a\in [a_{\rm min}, 5]$}.
\end{equation}
The lower boundary $a_{\rm min}$ is assumed to be a function of the stellar
radii of the primary and secondary star at the zero age main-sequence (ZAMS),
\begin{equation}
a_{\rm min}\equiv \frac{2 R_{\rm ZAMS,1} + 2 R_{\rm ZAMS,2}}{1-e}
\end{equation}
This was adopted to ensure that at zero age both stars are well
within their Roche lobes at closest approach, i.e.\ the periastron
distance $d_{\rm per}=a (1-e)$. The distribution of mass ratios is a 
pseudo-flat distribution. The only difference with the new
distribution concerns the boundaries.   \citet{Dominik+2012} consider $q 
\in [q_{\rm min},1]$ with $q_{\rm min}=0.08/m_1$ to avoid selecting
secondaries below H-burning limit. 
For the distribution of eccentricities a thermal-equilibrium
distribution is adopted \citep{Heggie1975}, which favors eccentric
systems:
\begin{equation}
f_e (e) \propto e, \quad \quad \text { for $e \in [0,1]$}.
\end{equation}

The binary fraction adopted by \citet{Dominik+2012} is $f_{\rm bin}=0.5$, 
corresponding to the parameter boundaries specified above.
For comparison we also provide estimates based on the old distribution but 
assuming a high binary fraction of $f_{\rm bin} = 1$ in model O-f1.

The distribution of primaries masses (and single star masses) is
identical to the one we adopt in the new reference model, given in
Eq~\ref{eq:imf}.

\section{Computational method}\label{sec:method}

For the purpose of this study we use the binary population synthesis
code  {\tt StarTrack} \citep{Belczynski+2002,Belczynski+2008}, which has
been used extensively to simulate the  evolutionary channels and
formation rates of compact objects {\tt www.syntheticuniverse.org}.  
The suite of our simulations is available on this website. 

This code belongs to a family of very fast and stable binary evolutionary codes, 
that provide powerful tools to explore the vast multi-dimensional space of the 
initial parameters that determine the fate of a binary system. These codes also 
enable the exploration of the effect of model uncertainties. These codes relay on  
precomputed grids of detailed stellar models  and approximate treatments of the 
physical processes as described below.  Despite the approximations, studies 
with this and similar codes based on the same philosophy have 
enabled insights into the many exotic phenomena resulting from low and high mass 
interacting binaries \citep[e.g. recent work by][]{Izzard+2004b, Izzard+2006, 
 Izzard+2009, Ruiter+2009, Mennekens+2010, Vanbeveren+2013, Fragos+2013, Abate+2013, de-Mink+2013, de-Mink+2014, Toonen+2014, Kochanek+2014, Claeys+2014, Mennekens+2014, Ruiter+2014, Schneider+2014, Schneider+2015}

A full description of the {\tt StarTrack} code used in this study and assumptions 
concerning the treatment of stellar evolution and binary interaction can be found 
in the papers above \citet{Belczynski+2002,Belczynski+2008}, \citet{Dominik+2012} 
and references therein. Below we provide a summary of the main assumptions relevant 
for this study.

\subsection {Physical assumptions} 

This study is a comparative study of the impact of the initial
conditions. \citet{Dominik+2012}  is used as our reference model.  We
use \citet{Dominik+2012}, also when the assumptions are uncertain.  

\paragraph{Stellar evolution} The code is based on detailed, but non rotating, single
stellar evolutionary models by \citet{Pols+1998}. For this purpose the
code utilizes  the fit formula to the detailed models by
\citet{Hurley+2000} with several adaptations described in
\citet{Belczynski+2002}. The  effects of mass loss and accretion are
simulated using algorithms originally developed  by \citep{Tout+1997,
Hurley+2002} to account for effects such as rejuvenation  of the
accreting star when appropriate.  

\paragraph{Stellar wind mass loss} We employ updated wind mass loss
rates, which include winds mass loss from early-type stars 
\citep{Vink+2001}, Wolf-Rayet stars \citep{Hamann+1998, Vink+2005} and
enhanced mass  loss rates for Luminous Blue Variables calibrated to
account for the observed distribution  of masses of known black holes
\citep{Belczynski+2010, Orosz+2011}.  As a result the  simulations allow
for the formation of black holes with masses up to 15\Msun~in a solar 
metallicity environment.  For sub-solar metallicity environments ($Z =
0.006$), black  holes up to 30\Msun~are produced, consistent with the
mass of the most massive known  stellar-origin black hole in the IC10 X-1 system
\citep{Prestwich+2007, Silverman+2008}. Under these  assumptions the
simulations predict the formation of black holes with masses up to
80\Msun~in low metallicity ($Z=0.0002$) environments for stars below initial
mass of $150\msun$.

\paragraph{Roche-lobe overflow and common envelope evolution} To
determine whether mass transfer through Roche-lobe overflow is stable we
consider the stellar type of the donor and mass ratio  as outlined in
\citep{Belczynski+2008}.  In the case of stable mass transfer, we 
assume that half of the mass lost by the donor is accreted by the
companion.  The  remainder is leaving the system carrying a specific
angular momentum of $2\pi a^2 /P$,  where $a$ and $P$ are the orbital
separation and period following \citet{Podsiadlowski+1992}.   Common
envelope evolution is accounted for using the classical
energy balance  formalism \citep{Webbink1984} adopting a value of
$\alpha_{\rm CE} =1$ for the envelope efficiency parameter. The
parameter $\lambda$ describing the binding energy of the envelope is
taken from fits by \citet{Xu+2010, Xu+2010a}, implemented as described
in \citet{Dominik+2012}. 

\paragraph{Remnant masses for compact objects} To obtain the final mass
of the compact  objects, we consider the formation of a proto-neutron
star and subsequent fall back of  material. These are inferred to be a
function of the carbon oxygen core mass as outlined in Section~4.2 of
\citet{Fryer+2012}. The employed scheme utilizes the rapid supernova model 
that is able to reproduce the apparent mass gap between NSs and BHs 
\citep{Belczynski+2012b}.
We also account for the formation of neutron stars
by electron capture supernovae  \citep[e.g.][]{Nomoto1984,
Podsiadlowski+2004}  as outlined in Section 4.4 of \citet{Fryer+2012}. 
For single stars this effectively results in the formation of neutron
stars for initial stellar masses above 7.6\Msun.   The transition from
neutron star to black hole is set at a mass of 2.5\Msun~for the
resulting compact object. For high metallicity ($Z=0.02$) this makes  
single stars above initial mass of $21\msun$ to form BHs.   

\paragraph{Supernova kicks} We account for the classical Boersma-Blaauw kick
\citep{Boersma1961,Blaauw1961} resulting from mass loss during the explosion. 
In addition, newly born compact objects receive natal kicks which are
given by a Maxwellian distribution with a 1D rms of $\sigma = 265 \kms$,
based on  the observed distributions of radio pulsars
\citep{Hobbs+2005}.   The kicks are lowered  proportional to the amount
of fall back as described in Section 4.5 of \citet{Fryer+2012}.   An
exception is made for neutron stars formed through electron capture
supernovae for  which no natal kicks are assumed in this simulation.  As
a result of this  the mergers of double neutron stars are
biased toward neutron stars formed through electron  capture supernovae.

\paragraph {Submodel A and B} To consider the large uncertainties concerning 
binaries that begin Roche-lobe overflow while the donor star crosses the  
Hertzsprung gap, we consider two extreme cases. In submodel A we consider all 
common envelope formation channels including donor stars in any post 
main sequence evolutionary stage. In submodel B we explicitly exclude all evolutionary 
channels where the  common envelope is initiated by a Hertzsprung gap donor star.  
Submodel B can be considered as a more conservative estimate of the merger rates 
for NS-NS, BH-NS, and BH-BH binaries \citep{Belczynski+2007a}.  A detailed 
study and revision of criteria for the onset of common envelope will be presented 
shortly  (Belczynski, Pavlovskii \& Ivanova, in prep.).

\paragraph{Metallicity} With the reach of advanced LIGO/Virgo mergers
will originate from all sorts of environments with high and low
metallicity.  We therefore consider two representative metallicities $Z=0.02$ 
(typical of solar neighborhood)  and $Z=0.002$ (typical of small starburst galaxies).  
The physically consistent way to derive merger rates would be to start from
an appropriate star formation history in the Universe, adopt a model of
metallicity evolution with redshift and then calculate  broad range of
population synthesis models with varying metallicity in order to obtain
current rate of mergers \citep{OShaughnessy+2008,OShaughnessy+2010,
Dominik+2013,Dominik+2015}.  Given the scope of this 
study we adopt a simpler approach and take an even mix of these two compositions 
as a crude  representation of metallicity distribution in local Universe
\citep[e.g.][]{Panter+2008}.  Although simplified, this is sufficient to judge
the overall impact of the change in initial conditions and their 
associated uncertainties.

\paragraph{Difference with respect to \citet{Dominik+2012} } 
Since the publication of the population synthesis calculations by \citet{Dominik+2012}, 
it was found that a technical bug was introduced during one of the annual code updates, 
related to the treatment of tidal locking. For detached binaries, in which both stars 
reside well within their Roche lobe, the effect of tides on the stellar and orbital 
spin is negligible. In such systems, the spins of the stars are typically not synchronized 
with the orbital motion. As the stars evolve they change their spin period, for example as 
a result of evolutionary expansion. At some point the spin and orbital period may become 
comparable. Such a system appears to be synchronized, at least momentarily, even though 
tides are not effective.   The simulations by \citet{Dominik+2012} incorrectly treated 
these systems as tidally locked, resulting in an incorrect further orbital evolution for 
specific cases. In general, this resulted in shrinking of the binary system with the 
overall effect of over predicting the double compact object merger rates. 

For majority of the cases the differences are small: within factor of $\sim 2$, i.e. 
comparable to differences arising from the change of initial conditions. This can be seen 
when comparing the results before the correction given in  Table~2 and~3 of \citet{Dominik+2012}, 
the standard model marked "S" in their paper and our results for the same assumptions, but 
after correction for the bug, as we provide in our Table~1, marked as the old standard model, "O". 

In two cases, both for solar metallicity submodels B,  involving binary merges with black 
holes the differences are larger. The BH-NS Galactic merger rate was revised from 
0.2~\perMyr \citep{Dominik+2012} to 0.06~\perMyr (current study). The BH-BH Galactic merger 
rate was revised from 1.9~\perMyr  \citep{Dominik+2012} to 0.22~\perMyr (current study).   

Although the changes for these high metallicity channels are substantial, we repeat that the overall rate is completely dominated by the low metallicity channels. Therefore the impact of this correction on the overall compact merger rates, masses and delay times presented by  \citep{Dominik+2012, Dominik+2013, Dominik+2015}
is negligible.  
 
\begin{figure*}[t]\center
\includegraphics[width=\textwidth]{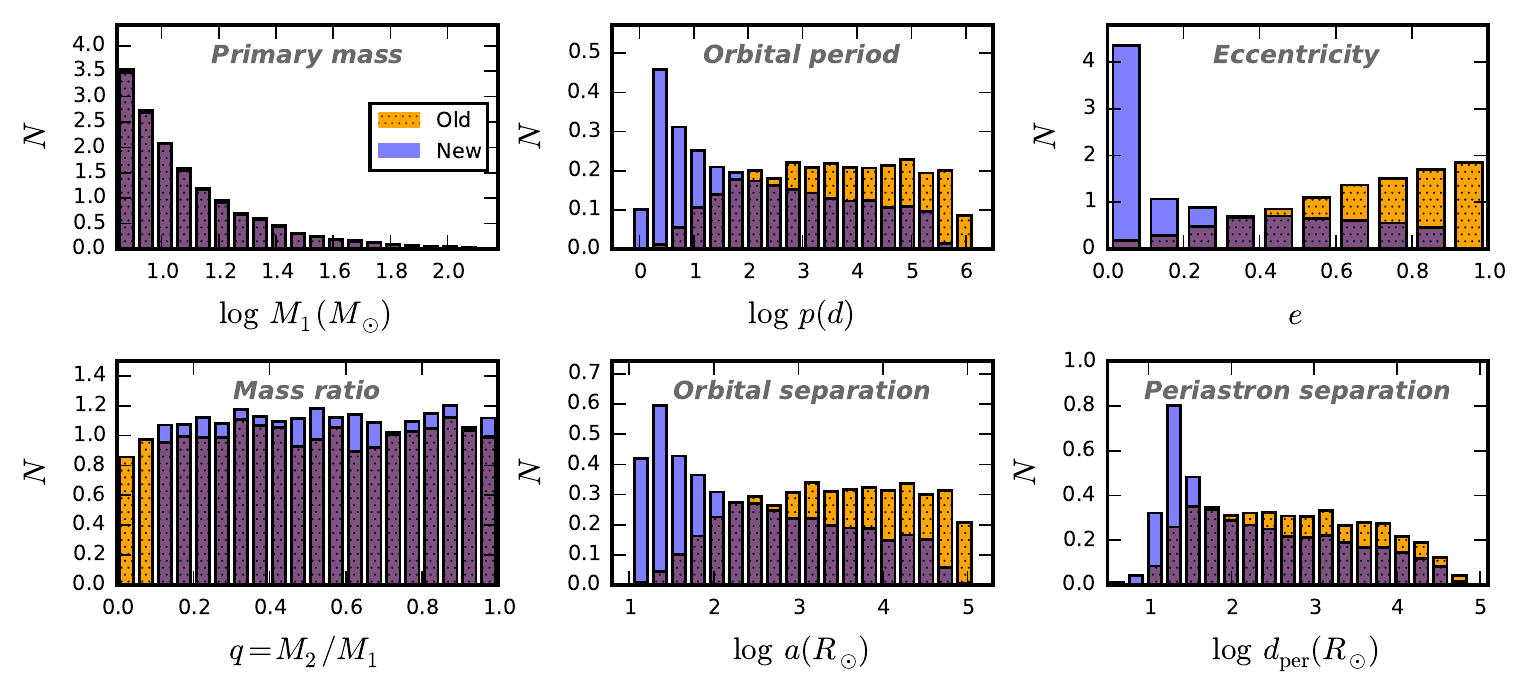}
  \caption{ Distributions of initial parameters for the simulated 
systems resulting from the old and new assumptions. All histograms are
normalized to unity. \label{initdist_all}}
\end{figure*}

\subsection {Computation of the merger rates}\label{sec:ratesmethod}

We randomly draw binary systems from the initial distribution functions, 
once for high metallicity ($Z=0.02$) and once for low metallicity  ($Z=0.002$). 
For computational efficiency we only select binary systems that are
massive enough, within some safety margin, to potentially produce a
double compact object involving a neutron star or a black hole. For
primaries we take  $M_1\ge5$\,\Msun~and for secondaries $M_2\ge3$\,\Msun. 

Using the {\tt StarTrack} population synthesis code we evolve $N_{\rm sim}
= 2\times 10^6$ of such binary systems. 
We estimate the mass formed in our simulations.  To obtain the 
total mass in stars down to the lower stellar mass limits we integrate over 
the full extent of the initial mass function ($0.08-150\msun$) and mass 
ratio distribution.  
 We include the appropriate mass for single stars when needed.  For this we assume that for each binary system with a primary mass $M_1$, there are $(1-f_{\rm bin})/f_{\rm bin}$ single stars with a mass $M = M_1$ 

We present our rates in terms of the rate for a fiducial Galaxy, ${\cal R}_{\rm gal}$. We chose to adopt the exact same method as \citet{Dominik+2012} to allow for comparison. 
 We therefore assume a $10$Gyr continuous star formation rate $\eta_{\rm SFR} = 3.5\Msun \,\myr^{-1}$. The resulting total mass formed in stars is within a factor of two of the estimates for the present-day stellar mass in the Milky Way\footnote{Estimates for the total stellar mass by \citet{Flynn+2006} yields $4.85 -5.5 \times 10^{10}\Msun$ and $6.43 \pm 0.63 \times 10^{10}\Msun$ by  \citet{McMillan2011}.}. 
The star formation history of our galaxy is not well known \citep[e.g.][]{Wyse2009} but there is evidence for several discrete epochs of enhanced of star formation \citep[e.g.][]{Cignoni+2006}. The assumption of continuous star formation may be a rather crude approximation for the Milky Way itself. It may however serve as the rate for a fiducial Milky Way-like galaxy obtained after averaging over the LIGO detection volume.

In Table~\ref{tab:Zdep} we provide the rates for the different simulations. 
These rates can be converted into approximate volumetric rates, as used
by the LIGO/Virgo collaboration, with the following expression.
\begin{equation}
{\cal R}_{\rm vol} =  10 \ {\rm yr}^{-1} \ {\rm \Gpc}^{-3}\ 
\left[	
	\frac{\rho}{0.01 \Mpc^{-3}} 
\right]
\left[	
	\frac{ {\cal R}_{\rm gal}}{ \myr^{-1} } 
\right]	
\end{equation}
where $\rho$ is the local density of Milky Way like galaxies and $R_{\rm gal}$ the 
Galactic merger rate.  

The detailed choices on how to normalize the rates are, to
some extent, arbitrary. This is a result of the uncertainties in
the low mass binary fraction and the poorly constrained distribution of
their mass ratios. We chose to adopt very simple assumptions and we assume
for the sake of this study that both binary fraction and mass ratio
distribution adopted in a given model for massive stars hold true for the 
entire considered mass range ($0.08-150\msun$). 

Our Monte Carlo simulations are subject to statistical fluctuations due to they 
finite size.  We estimate this by counting the total number of double compact 
objects that merge with 10 Gyr in each simulation, $N_{\rm x}$, considering NS-NS, 
BH-NS, BH-BH separately. We use $1/\sqrt {N_x}$ as an approximation for the 
statistical uncertainty.

\section { Results: distributions} \label{sec:dist}
\subsection { Comparison of the initial distributions
\label{sec:results:initdist} }
The effective initial distributions for all systems with a primary mass of at 
least 7\Msun~are shown in  Fig.~\ref{initdist_all} for the old and new assumptions.  
Systems with lower primary masses are not capable of producing double neutron stars 
in our simulations. The histograms are shown for a metallicity $Z=0.02$ and 
normalized to unity. The main differences between the old and new assumption 
concern (a) the period/separation distribution and (b) the eccentricity distribution. 

\paragraph{Periods and separations} The main difference between old and new 
assumptions concerns the tightest systems. The new distributions show a clear 
preference for short periods / small separations as can be seen in central top and 
bottom panels of Fig.~\ref{initdist_all}.  The old distribution is flat for large 
separations but tight systems ($a \lesssim 100 \Rsun$ ) are strongly suppressed. 
This is a result of the adopted lower limit $a_{\min}$, which depends on  the 
eccentricity and the stellar radii ( see Sect.~\ref{sec:initdist}).  Such a turnover 
is not observed for massive binaries.  Even though we do not know at present how such 
close systems form, 
they exist and must be accounted for in the simulations as we do in the new initial 
conditions. 

\paragraph{Eccentricities} The second main
difference concerns the eccentricities  (depicted in the top right panel of
Fig.~\ref{initdist_all}).  The distribution adopted in the old
simulations is the thermal distribution which strongly favors eccentric systems. 
The new distribution favors (near) circular systems.  The high
eccentricities in the old simulations allow much wider systems to still
interact as discussed below.  

\paragraph{Periastron separations}

The distribution of periastron separations $d_{\rm per} = a(1-e)$, i.e. the 
distance of closest approach at zero age, is shown in the bottom right panel 
in Fig.~\ref{initdist_all}.   To first order this is the most relevant parameter 
governing when systems
start to interact by tides and later mass transfer.  The old simulations
favor wider and more eccentric systems, while the new simulations favor
tighter and more circular systems.  By coincidence,  these changes partially  
compensate each other. 

The old assumptions resulted in a nearly flat distribution of periastron 
separations extending from $\log d_{\rm per} (\rsun) = 1$-$5$, being nearly 
flat between $\log d_{\rm per} (\rsun) = 1.5$-$4$ and turning over on both ends.  
Instead the new distributions peak at short periastron separations of about 
$10$-$50\rsun$ and gradually decay for higher separations.  

However, in the regime of relevance for the formation of double compact objects 
($\log d_{\rm per} (\rsun) = 1.5$-$4$, as we will see in the next section)
both, old and new, periastron distribution are rather similar. 
We note only a slight domination of old distribution in this regime.  
The distributions in this figure are normalized to unity. The larger overall 
binary fraction in the new simulations will compensate for this. 

\paragraph{Primary masses and mass ratios}  In both simulations the mass
ratios are drawn from a flat seed distribution (the bottom left panel
of Fig.~\ref{initdist_all}). They only differ in the adopted boundaries.
In the old simulation companion masses are drawn down to the hydrogen
burning limit.  In the new simulations the minimum mass ratio adopted is
${q_{\rm min} =  0.1}$.  In our simulations with old initial
distributions we found no double compact objects resulting from systems
with such extreme mass ratio. This difference therefore only has a very minor 
effect on the final normalization. 

The distributions of initial primary masses $M_1$ (shown in the top left
panel of Fig.~\ref{initdist_all}) are identical in both simulations. 
The only differences are stochastical in nature.

\subsection {Comparison of the birth properties of the progenitors of double 
compact object mergers \label{sec:progdist}}
\begin{figure*}[t]\center
 \includegraphics[width=\textwidth]{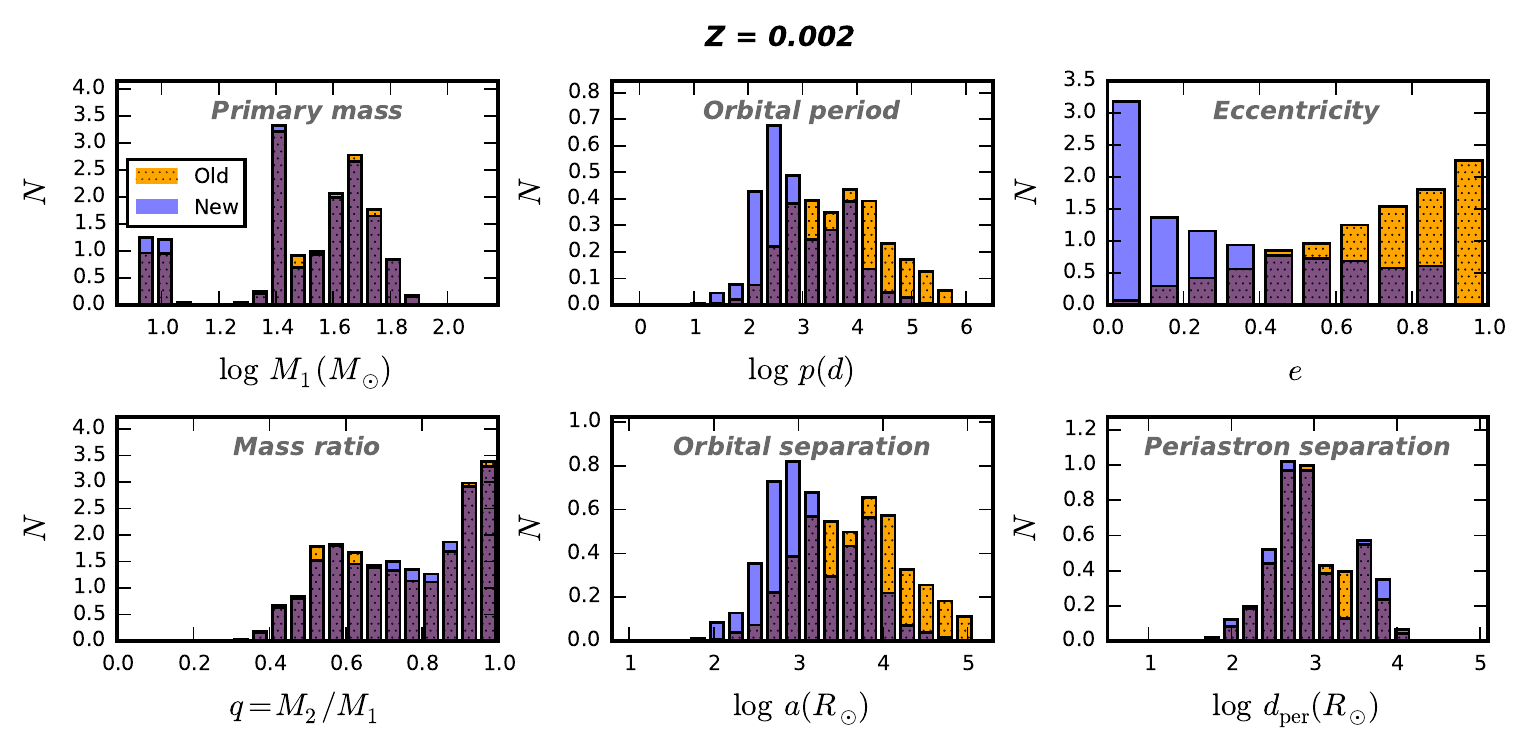}  
 \includegraphics[width=\textwidth]{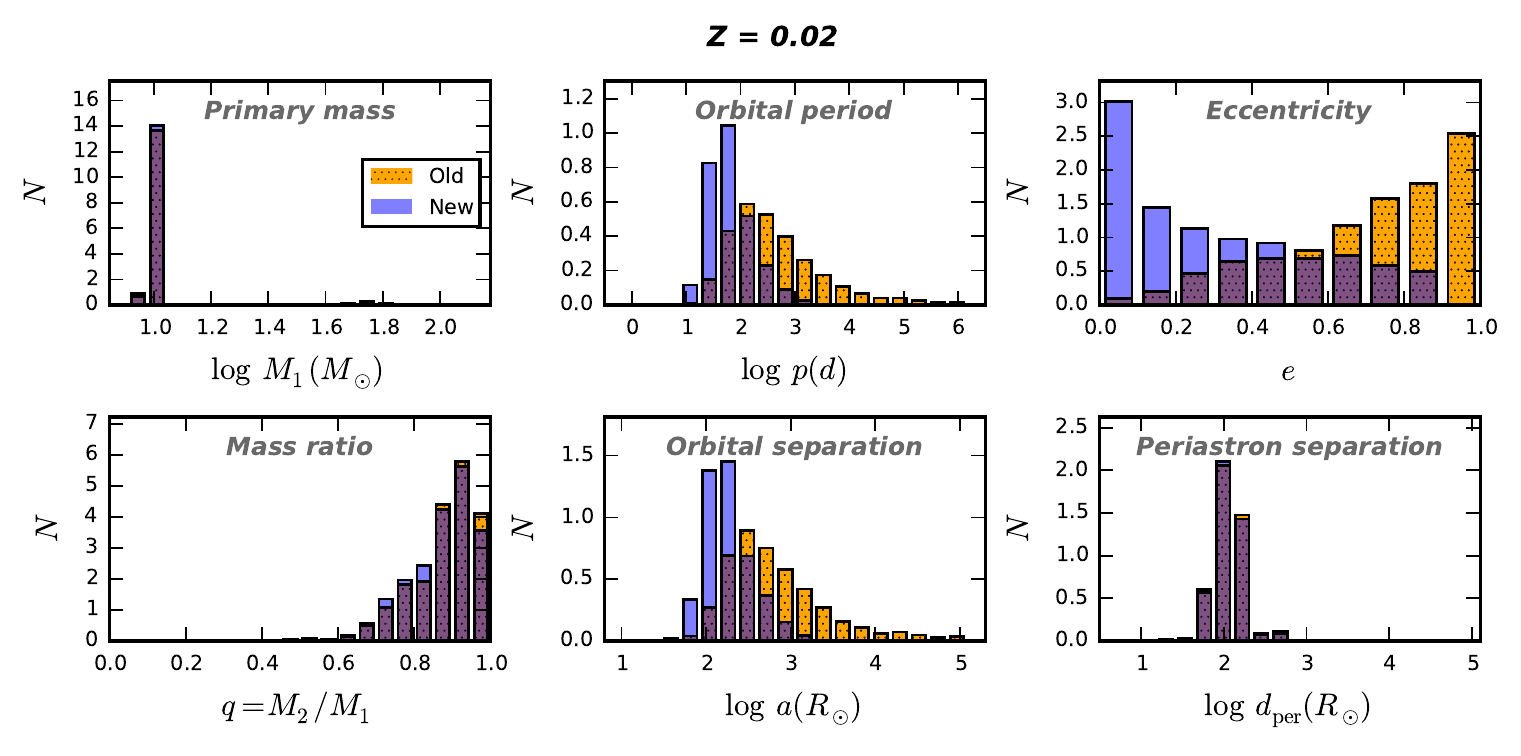}  
  \caption{ Birth distributions of initial binary parameters for the
progenitors of double compact object mergers. The normalized histograms
show the distributions for submodel B combining all types of mergers,
for low and high metallicity.  \label{initdist_progenitors}}
\end{figure*}

\begin{figure*}[t]\center
\includegraphics[width=0.65\textwidth]{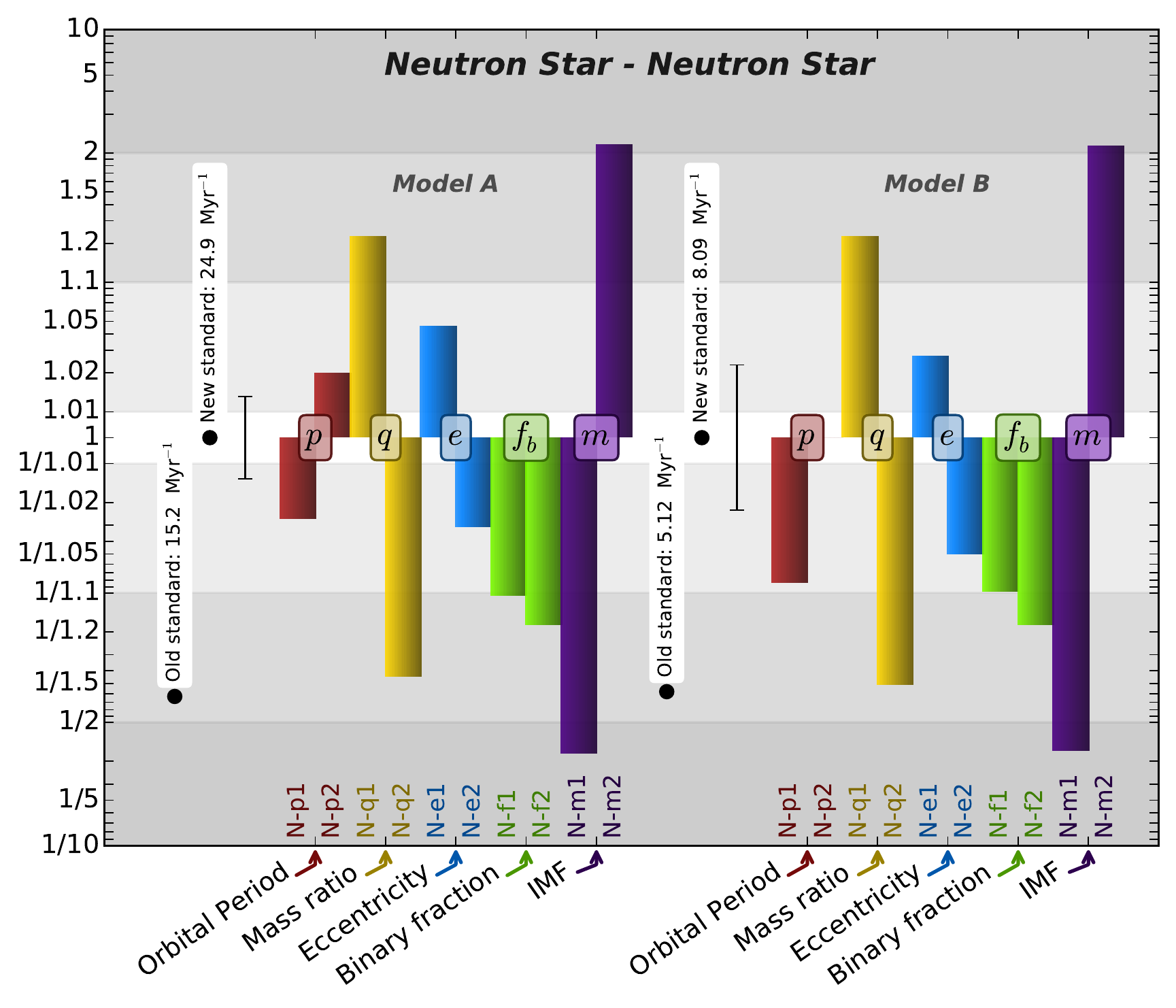}
  \caption{ Variations in the merger rate predictions double neutron stars for the old and
new standard assumptions and the effect of the allowed variations in the
initial distributions of binary parameters. See also
Table~2.  Note that the Poisson variations
(indicated as error bars) resulting from the finite size the number of
simulated systems are comparable with some of the variations.  We use a
double mirrored logarithmic vertical axis to visualize the very small
variations we obtain in most cases. For comparison, uncertainties from
other sources (e.g., evolutionary uncertainties) are typically an order 
of magnitude or more. 
  \label{fig:uncertainties}}
\end{figure*}

\begin{figure*}[t]\center
\includegraphics[width=0.5\textwidth]{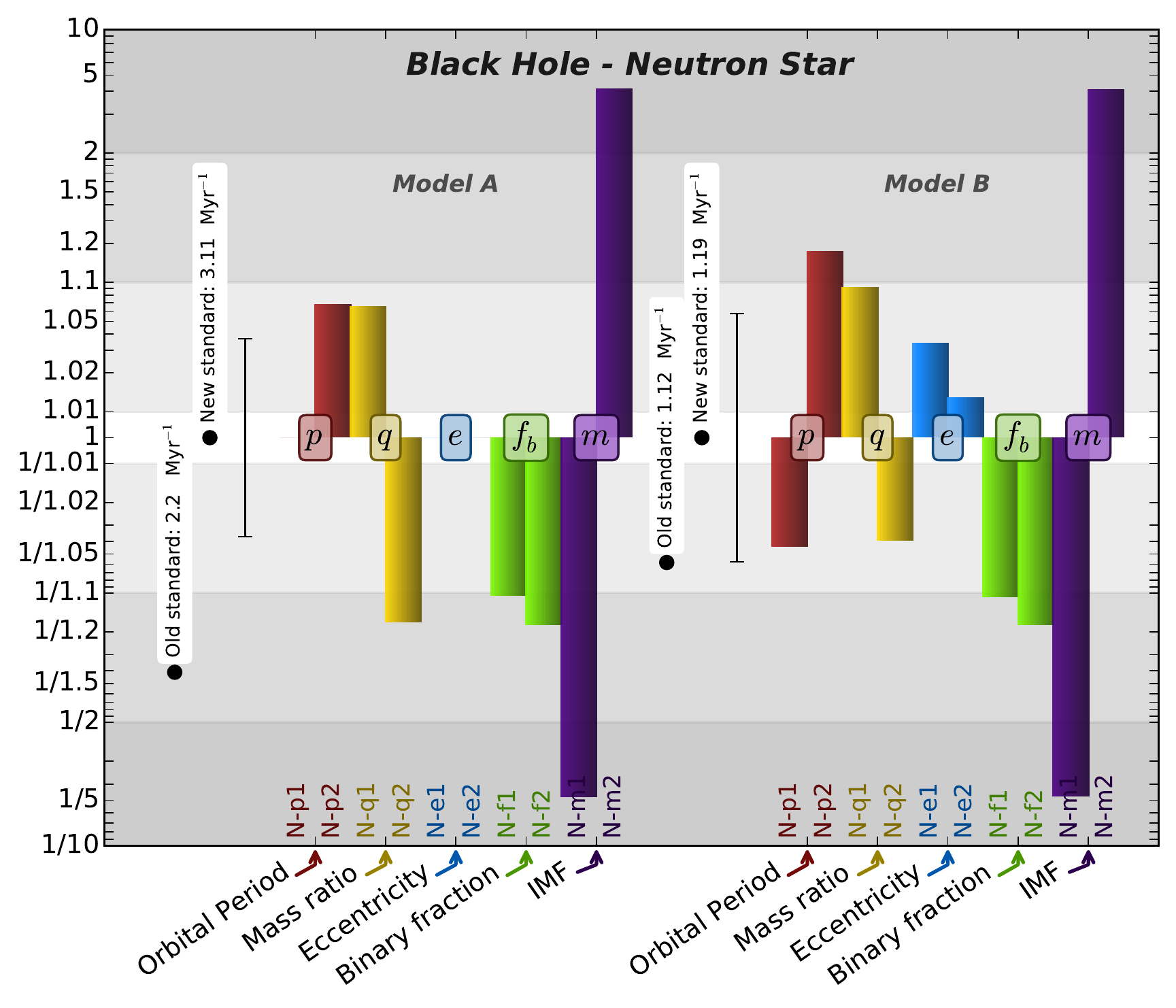}%
\includegraphics[width=0.5\textwidth]{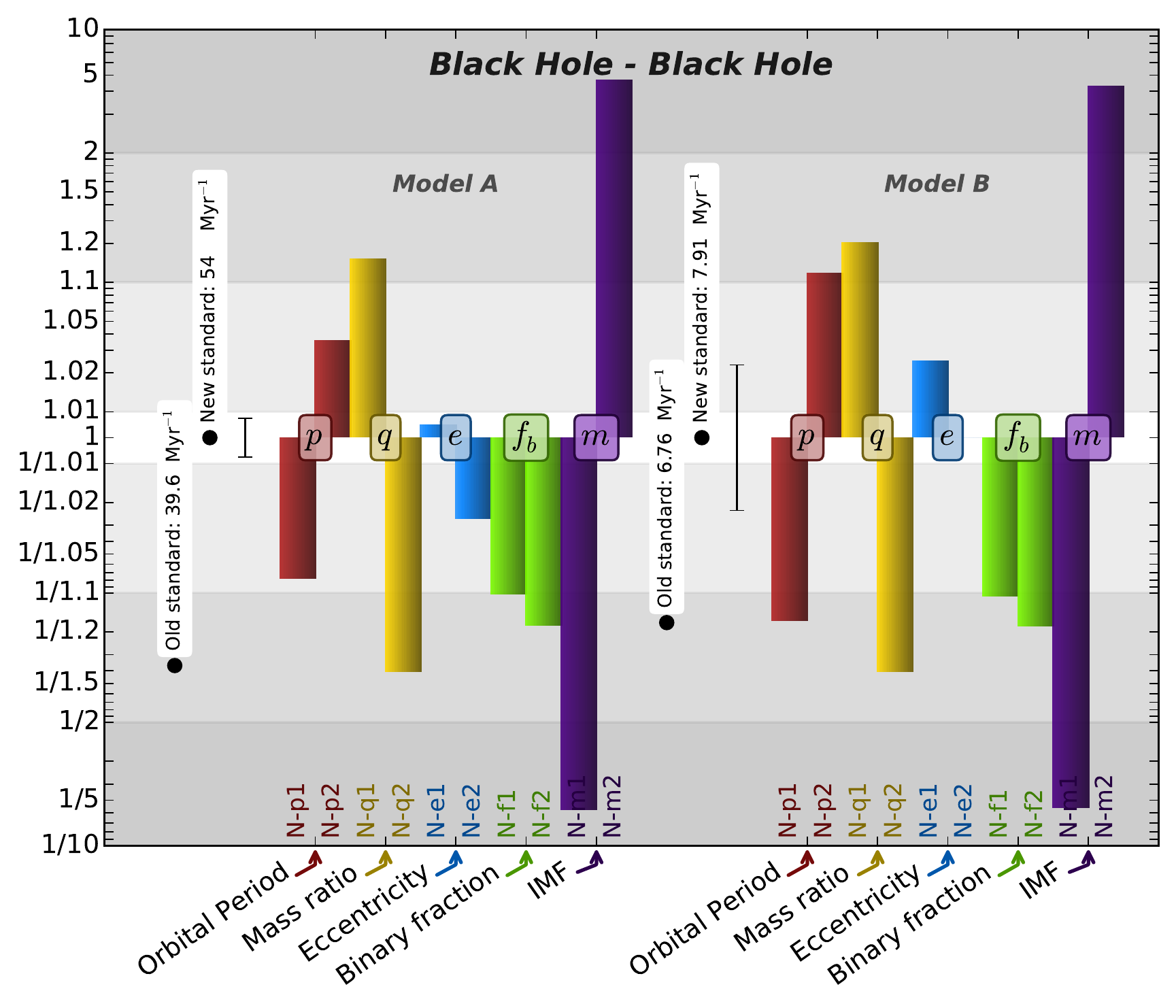}
  \caption{ As Fig.~\ref{fig:uncertainties}  for the merger rates BH-NS (left) and BH-BH (right) systems. 
  \label{fig:uncertainties-BH}}
\end{figure*}

Only a very small subset of the simulated binaries produces two compact
objects that are bound and that will merge within 10\,Gyr. The birth properties 
of this subset are shown in Fig.~\ref{initdist_progenitors}, where we compare 
the impact of the old and new assumptions for the initial conditions. 

Note that the histograms are normalized to unity to allow comparison of the 
shape of the distribution, the resulting rates are discussed below. We separately show results for low metallicity (Z = 0.002, top panels) and high metallicity (Z = 0.02, bottom panels).

Since the trends observed when comparing the old and new initial distributions are very similar 
for submodel A and B, we chose to only show the results submodel B here. This is the more conservative submodel which excludes progenitor systems  in which a Hertzsprung gap star acts as donor star during CE phase, see Section~3.1.  In practice this means that on average the progenitors in model B result from wider systems and show a slightly stronger preference for equal mass progenitor systems.  Data for submodel A (as well as data subdivided by the double compact object type) can be obtained from the online data repository for readers that are interested. More insight in the differences between submodel A and B can be found in \citep{Dominik+2012}.

\paragraph{Primary masses} 
The distribution of initial primary masses of the progenitors consists
of separate components. The narrow low mass peak around $10\msun$ is 
generated by the progenitors of double neutron stars. The high mass component 
is only prominent in the low metallicity simulations. This component
consists of the progenitors of mergers that involve a black hole. 

The minimum initial mass to form a black hole is  $40\msun$ for the primary 
star of a  double compact object progenitor at  $Z=0.02$.  This is much 
higher than the minimum mass for a single star (or star in a very wide non 
interacting binary) to form a black hole in our simulations, which is  
$21\msun$ at  $Z=0.02$. The difference is the result of the additional mass 
loss experienced in binary systems due to Roche lobe overflow and common 
envelope ejection from BH progenitor.  At lower metallicity these minimum 
masses decrease. At  $Z=0.002$ we find a minimum initial  mass of about 
$20\msun$ for the primary star of a double compact object progenitor 
involving a black hole.  

The difference between the old and new simulations is very small. At low 
metallicity there is a slight tendency towards lower mass progenitors 
(favoring NS-NS progenitors) in the new simulations.  

\paragraph{Mass ratios}
For both the old and new simulations we find that the progenitors
exclusively result from systems with mass ratios larger than about
$M_2/M_1 \gtrsim 0.4$ at low metallicity and about 0.6 at high metallicity. 
For both metallicities there is a preference for progenitors with very
equal masses, i.e.  0.9 and higher. 

The difference between the old and new simulations is marginal,
unsurprisingly given the similarity of the input distribution.  A very
slight tendency for the old simulations to favor more equal mass systems
is observed at high metallicity, while the opposite behavior is observed 
at low metallicity. However, the differences are insignificant in comparison 
with the model uncertainties.

\paragraph{Periods, separations and eccentricities}
The initial period and separation distributions of the progenitor
systems show a clear difference between the old and new initial conditions. 
For both metallicities the old simulations are biased towards wider systems
than the new simulations. At high metallicity the distributions peak near 
$150\rsun$ in the new simulations and near  $300\rsun$ in the old simulations. 
At low metallicity the distribution is
clearly bimodal peaking around $800$ and $8000\rsun$ in the new simulations. 

The distribution of initial eccentricities of merger progenitors closely 
reflects the shape of the input distribution of the population, although
in both cases systems with larger eccentricities are over represented. 

\paragraph{Periastron separations}
Although the distributions of initial separations $a$ and eccentricities
$e$  are very different between the old and new simulations, the
distribution of initial periastron separations $d_{\rm per}$
of merger progenitors is nearly indistinguishable. 

For low metallicity the distribution is again bimodal.  Detailed inspection 
of the evolutionary paths of the systems that populate both peaks show that 
the tighter systems first evolve through a
stable mass transfer phase. The wider systems first evolve through a
common envelope phase. These evolutionary channels are identified and
described in earlier work (\cite{Dominik+2012}; see their Table 5 and
associated text). 

Both channels exist in the old and new simulations, since the input physics  
assumptions are not changed.  The similarity in the periastron separation 
distributions implies that the relative contribution of both channels did not 
significantly change when switching from the old to new assumptions.

\subsection { Distributions of final properties }

The distribution of the final properties predicted by our simulations that are 
of interest for gravitational wave searches include the delay times, component 
masses, mass ratios and chirp masses. Detailed models of the anticipated waveforms 
will enable the inference on several parameters, such as component masses 
\citep{Aasi+2013b, Veitch+2015} and if multiple events are detected constraints on the 
properties of the population can be inferred \citep{Mandel2010, Mandel+2015}.

We find that the distributions of final properties are remarkably insensitive to 
the assumed initial conditions.  The predicted distribution of the component masses, 
mass ratios, chirp masses and delay times show no significant changes when the 
initial conditions are varied within current uncertainties.  They are almost 
identical to those presented by \cite{Dominik+2012}, to which we refer for detailed 
discussion.   

The insensitivity of rate predictions results from the (coincidental) 
similarity of the initial periastron separation distribution and the mass ratio 
distribution, which remained practically unchanged. These two parameters are key in 
determining the fate of the binary system (and thus merger rates). However, the 
physical properties of merging binaries are not sensitive to changes in initial 
conditions. 

There are relatively few formation channels for double compact 
object mergers. For a given model of evolution, these formation channels originate 
from a rather narrow volume of initial parameter space (``the formation volume''). 
Change of initial conditions alters the number of the progenitor binaries in the 
formation volume, but has very little effect on the binary merger properties.
This insensitivity is not just connected to the specific initial conditions 
considered here, but is of more fundamental nature.  It is a robust result that has also been 
found in connection to other compact object binaries. For example by \citet{Kalogera+1998} in the context of low-mass X-ray binaries, and see also \citet{Belczynski+2002}. 

The robustness of the final distributions against initial condition uncertainties considered 
here, but also wider variations,  is reassuring in the light of anticipated gravitational wave 
detections. It removes one layer of uncertainties and allows for more optimism concerning the 
potential of such detections for constraining more interesting aspects such as the physics 
uncertainties.

The final properties of our simulations are available through our online data files  available 
at ({\tt www.syntheticuniverse.org}).

\section{Results: Merger Rates} \label{sec:rates}

In Table~\ref{tab:Zdep} we list the double compact object merger rates for our
new standard model N, the old standard model O,  and variations on the
models that explore the effect of uncertainties in initial distributions.  
All rates are expressed as the rate for a fiducial Milky Way Galaxy as detailed in 
Sect~\ref{sec:ratesmethod}. We give the results both for submodel A and 
the more conservative submodel B. Merger rates are also presented 
separately for low ($Z=0.002$) and high metallicity ($Z=0.02$) stellar and 
binary evolution. 

In Table~\ref{tab:Mix} we present merger rates that correspond to a 
fiducial Galaxy with an even mix ($50\%$--$50\%$) of low-metallicity
($Z=0.002$) and high-metallicity ($Z=0.02$) stars.  Absolute mixed merger 
rates are given only for our new standard model (an average of low- and 
high-metallicity fiducial Galactic rates from Table~\ref{tab:Zdep}) and 
can be readily obtained for any other model. For all other models only 
relative change of rates with respect to our new standard is given.
A graphical representation of the changes relative to the new standard
model is given in Figure~\ref{fig:uncertainties}.

The overall conclusions are: {\em (i)} merger rates increase
slightly with the new assumptions, and they are within 
a factor of $2$ of old standard rates; 
{\em (ii)} merger rates change by
less than factor of $2$ within associated uncertainties embedded in new initial
distributions concerning binary parameters (the only exception being the IMF). 
These conclusions hold for all types of double compact object mergers
(NS-NS, BH-NS and BH-BH) and the changes are typically much smaller
than the maxima listed above. In the following section we discuss the various
trends.

\subsection {Impact of the new initial conditions}

The adoption of the new assumptions for the initial conditions leads to higher 
estimates for all merger rates.  The largest change is found when adopting 
model A. The old rates are  39\% lower than the new rates for  NS-NS mergers, 
29\% lower for BH-NS mergers and 27\% lower for 
BH-BH mergers. An overview of the relative changes is given in 
Tab.~\ref{tab:Mix} using the new standard rate as reference.   

When adopting submodel B the observed changes are even smaller in the case of 
BH-NS and BH-BH mergers.  The submodel B excludes progenitor channels 
which involve common envelope evolution with a Hertzsprung gap donor \citep{Belczynski+2007a}. These 
typically come from progenitor binaries with smaller initial separations where 
the old and new distributions deviate most strongly. Excluding these
progenitors reduces the difference between old and new simulations.  
For the BH-NS in submodel B the changes are so small that they are of 
the same order as the expected statistical fluctuations resulting from 
the finite number of systems in our simulations. These average statistical 
fluctuations are shown in Figure~\ref{fig:uncertainties} as an errorbars.  

The main reason for the slight increase in the merger rate is the large  
binary fraction adopted in the new initial conditions ($f_b=1$). The old distributions 
with the high binary fraction of $f_b =1$ (model O-f1) give merger rates that 
are higher (up to $\sim 50\%$ in the case of BH-NS mergers in submodel B) than for 
the new standard model (see Table 2).

\subsection {Variations due to uncertainties in the new initial conditions}
All relative merger rate changes are smaller than $34\%$ for {\em all}
types of double compact objects when varying the initial conditions
concerning binary properties (models N-p1, N-p2, N-q1, N-q2, N-e1, N-e2, 
N-f1, N-f2; see Tab.~\ref{tab:Mix}).   This is not true for variations 
in the IMF, which we will discuss at the end of this paragraph. 

The initial distribution of mass ratios that rather steeply falls off with 
mass ratio (model N-q2) is the cause of merger rate decrease of the order of 
$\sim 30\%$ for NS-NS and BH-BH mergers.  
A distribution favoring equal mass systems at birth mildly favors the
formation of NS-NS systems and BH-BH systems. The BH-NS mergers show the 
same trend, but the changes  are comparable to the Poison fluctuations.  
Varying the initial period distribution has a smaller effect, below $20\%$ 
and approaching our statistical uncertainties. A flatter period distributions, 
i.e. one that favors wider systems (model N-p2), results in a more effective 
formation of compact object mergers. A period distribution favoring short 
period systems more strongly (model N-p1) mainly adds systems that 
interact or merge prematurely, preventing the formation of a double compact 
object system.
Varying the eccentricity distribution within the ranges given by 
\citet{Sana+2012}  has no significant effect on the merger rates.  All 
variations found are less than $5\%$ for all submodels and all types of 
mergers considered. These variations are dominated by the statistical 
fluctuations in our Monte Carlo simulations.  
A reduction of the total binary fraction to $f_{\rm bin} = 0.85$ \citep[which roughly corresponds to the one sigma lower limit of intrinsic spectroscopic binary fraction, $f_{\rm sp} = 0.7 \pm 0.1$, derived by ][]{Sana+2012} leads to a 
reduction of all compact object merger rates by $10\%$ (model N-f1).  A further 
reduction of the total binary fraction to $f_{\rm bin} = 0.7$ (model N-f2), 
which roughly corresponds two the two sigma lower limit on the observational constraints, 
reduces the merger rates by $15\%$.

Significant change of the power law exponent of the IMF for stars more 
massive than $1\msun$ generates the largest change  of double compact 
merger rates in our suite of models. We allow for a change of the exponent by 
$0.5$ up and down from our standard value $\alpha=2.7$. Note that this affects 
the mass distribution of the primary stars in binary systems and single stars 
for those models where $f_{\rm bin} < 1$ (model O-f1, N-f1, N-f2).
Merger rates decrease within a factor of $2.7$ for NS-NS mergers, a factor of
$4.8$ for BH-NS mergers and a factor of $5.8$ for BH-BH mergers for our most
steep IMF (model N-m1) in respect to our new standard model (N).
Merger rates increase within a factor of $2.2$ for NS-NS mergers, a factor of
$4.1$ for BH-NS mergers and a factor of $4.7$ for BH-BH mergers for our most
flat IMF (model N-m2).
These changes are due to two effects. First, the steeper the IMF the fewer double 
compact objects form in a population of binary stars. The heaviest BH-BH
mergers are affected the most. Second, the slope of high-mass IMF has a 
significant impact on the total mass of the fiducial Galaxy. For the same
number of massive binary stars that we draw from initial distributions 
($N_{\rm sim}=2 \times 10^6$) the corresponding total mass of simulation (over 
the entire stellar mass range) is $\sim 2.5$ higher and $\sim 2.5$ smaller for 
our model with the steep and flat IMF, respectively as compared with our
adopted standard IMF.

\begin{deluxetable*}{l cc | cc | cc  c cc | cc | cc   r}
\tabletypesize{\small}
\tablecaption{Merger rates for a fiducial Milky Way Galaxy [Myr$^{-1}$]} 
\tablehead{
 & \multicolumn{6}{c}{{\bf Z = 0.002} } & & \multicolumn{6}{c}{{\bf Z = 0.02} } & \\
	Model ID & 
	\multicolumn{2}{c}{NS-NS} & 
	\multicolumn{2}{|c|}{BH-NS} & 
	\multicolumn{2}{c}{BH-BH} & &	
	\multicolumn{2}{c}{NS-NS} & 
	\multicolumn{2}{|c|}{BH-NS} & 
	\multicolumn{2}{c}{BH-BH} & 
	Description \\
 & {A} &  {B} & {A} &  {B} & {A}& {B} & &
 {A} &  {B} & {A} &  {B} & {A}& {B}  & }
\startdata

&&&&& &&&&&&&\\

N     &   12.8 &   3.68 &   4.45 &   2.32 &   98.3 &   15.6 &    &  36.9 &   12.5 &   1.77 &   0.06 &   9.73 &   0.22 & new standard \\

&&&&& &&&&&&&\\
O     &   8.17 &   2.43 &   3.17 &   2.24 &   72.2 &   13.3 &&  22.3 &   7.82 &   1.24 &   0.01 &   6.98 &   0.22 & old standard \\
O-f1  &   13.2 &   3.92 &   5.11 &   3.60 &   116  &   21.5 &&  35.9 &   12.6 &   1.99 &   0.02 &   11.3 &   0.36 & old, $f_{\rm bin}=1.0$ \\
&&&&& &&&&&&&\\
N-p1  &   12.5 &   3.34 &   4.72 &   2.23 &   90.8 &   13.3 && 35.9 &   11.6 &   1.46 &   0.05 &   9.46 &   0.29 & new, $(\log p)^{-0.75}$ \\
N-p2  &   13.4 &   3.67 &   4.64 &   2.75 &   102  &   17.5 &&  37.3 &   12.6 &   2.00 &   0.05 &   9.87 &   0.20 & new, $(\log p)^{-0.35}$ \\
&&&&& &&&&&&&\\
N-q1  &   16.3 &   4.28 &   5.15 &   2.53 &   113  &   18.8 && 44.7 &   15.6 &   1.48 &   0.07 &   11.7 &   0.24 & new, $q^{0.5}$ \\
N-q2  &   8.65 &   2.52 &   4.00 &   2.23 &   69.8 &   11.1 &&  25.7 &   8.17 &   1.32 &   0.06 &   6.84 &   0.13 & new, $q^{-0.7}$ \\
&&&&& &&&&&&&\\
N-e1  &   13.2 &   3.71 &   4.49 &   2.40 &   99.0 &   16.0 && 38.8 &   12.9 &   1.70 &   0.06 &   9.93 &   0.22 & new, $e^{-0.59}$ \\
N-e2  &   12.7 &   3.41 &   4.66 &   2.37 &   95.3 &   15.7 && 35.5 &   12.0 &   1.57 &   0.04 &   9.90 &   0.12 & new, $e^{-0.25}$ \\
&&&&& &&&&&&&\\
N-f1  &   11.6 &   3.34 &   4.03 &   2.09 &   89.1 &   14.1 && 33.4 &   11.4 &   1.60 &   0.06 &   8.84 &   0.20 & new, $f_{\rm bin}=0.85$ \\
N-f2  &   10.9 &   3.13 &   3.78 &   1.97 &   83.4 &   13.2 && 31.3 &   10.6 &   1.51 &   0.05 &   8.26 &   0.19 & new, $f_{\rm bin}=0.7$ \\
&&&&& &&&&&&&\\
N-m1  &   4.70 &   1.43 &   0.99 &   0.49 &   17.2 &   2.77 && 13.4 &   4.65 &   0.31 &   0.01 &   1.54 &   0.04 & new, ${m_1}^{-3.2}$ \\
N-m2  &   26.5 &   7.80 &   17.9 &   9.51 &   457  &   66.9 && 81.3 &   26.9 &   7.86 &   0.32 &   51.4 &   1.19 & new, ${m_1}^{-2.2}$ \\
\enddata
\label{tab:Zdep}
\tablecomments{$^{\rm a}$Rates correspond to a galaxy with $10$~Gyr of constant star
formation rate at the level of $3.5\mpy$ as detailed in Section~3.2. Rates are given 
for low and high metallicity and for two different model assumptions. In model A we
allow for high rate of common envelope survival, while more the conservative model B does
not allow the formation of double compact objects for donors on Hertzsprung 
gap.}
\end{deluxetable*}

\begin{deluxetable*}{l  cc | cc | cc  | c r}
\tabletypesize{\small}
\tablecaption{Relative Changes of Mixed-Metallicity Merger Rates$^{\rm a}$} 
\tablehead{Model ID & \multicolumn{2}{c}{NS-NS} &
\multicolumn{2}{|c|}{BH-NS} & \multicolumn{2}{c}{BH-BH} & Description \\
 & {A} &  {B} & {A} &  {B} & {A}& {B} &  &}
\startdata
N     & 1.000   & 1.000   & 1.000   & 1.000   & 1.000   & 1.000   &  new standard \\
      & {\bf(24.85)} & {\bf(8.090)} & {\bf(3.110)} & {\bf(1.190)} &  {\bf(54.02)} & {\bf(7.910)} & { absolute rates [Myr$^{-1}$]}\\
&&&&&&\\
O     & 0.613   & 0.633   & 0.709   & 0.945   & 0.733   & 0.855   &  old standard \\
O-f1  & 0.988   & 1.021   & 1.141   & 1.521   & 1.178   & 1.382   &  old, $f_{\rm bin}=1.0$ \\
&&&&&&\\
N-p1  & 0.974   & 0.923   & 0.994   & 0.958   & 0.928   & 0.859   &  new, $(\log p)^{-0.75}$ \\
N-p2  & 1.020   & 1.006   & 1.068   & 1.176   & 1.036   & 1.119   &  new, $(\log p)^{-0.35}$ \\
&&&&&&\\
N-q1  & 1.227   & 1.229   & 1.066   & 1.092   & 1.154   & 1.204   &  new, $q^{0.5}$ \\
N-q2  & 0.691   & 0.661   & 0.855   & 0.962   & 0.709   & 0.710   &  new, $q^{-0.7}$ \\
&&&&&&\\
N-e1  & 1.046   & 1.027   & 0.995   & 1.034   & 1.008   & 1.025   &  new, $e^{-0.59}$ \\
N-e2  & 0.970   & 0.952   & 1.002   & 1.013   & 0.974   & 1.000   &  new, $e^{-0.25}$ \\
&&&&&&\\
N-f1  & 0.905   & 0.911   & 0.905   & 0.903   & 0.907   & 0.904   &  new, $f_{\rm bin}=0.85$ \\
N-f2  & 0.849   & 0.849   & 0.850   & 0.849   & 0.848   & 0.846   &  new, $f_{\rm bin}=0.7$ \\
&&&&&&\\
N-m1  & 0.364   & 0.376   & 0.209   & 0.210   & 0.173   & 0.178   &  new, ${m_1}^{-3.2}$ \\
N-m2  & 2.169   & 2.145   & 4.141   & 4.130   & 4.706   & 4.304   &  new, ${m_1}^{-2.2}$ \\
\enddata
\label{tab:Mix}
\tablecomments{Rates correspond to a fiducial Milky Way Galaxy with an even mix 
($50\%$--$50\%$) of 
low-metallicity ($Z=0.002$) and high-metallicity ($Z=0.02$) stars (the average of low- and 
high-metallicity fiducial Galactic rates from Table~\ref{tab:Zdep}). We give all rates 
relative to our new standard model.  A graphical representation is shown in 
Figure~\ref{fig:uncertainties}.}
\end{deluxetable*}

\section{Discussion and Conclusion} \label{sec:concl}

Recently, \cite{Sana+2012} have presented measurements of initial
binary parameters for massive stars. These measurements, in some aspects, 
are in stark contrast with so far used assumptions in evolutionary studies 
of massive stars \citep[e.g.][]{Belczynski+2002,Voss+2003,Dominik+2012,Mennekens+2014}. The old standard prescriptions 
(a flat distribution in log separation,  a thermal eccentricity distribution 
strongly favoring eccentric systems and typically an adopted binary fraction of 
50\% counting systems with separations up to $10^5$\Rsun) may still apply for low- 
and intermediate-mass stars, but the new 
measurements show very different initial binary parameter distributions for 
massive stars (strong overabundance of very tight systems, a binary fraction 
that reaches $100\%$ when including systems up to about $10^{4.5}\rsun$  and 
mostly circular systems). 

We find a small increase in the NS-NS, BH-NS and BH-BH merger rate as a result 
of the new initial conditions (due to the increased binary fraction), with 
the old rates being up to $40$\% lower than the new rates. Secondly, we study 
the impact of the allowed variations of the 
new binary parameter distributions within the observational uncertainties. We 
find that they have negligible impact (typical variations of a few percent, 
always less than a factor of $2$).  All changes are negligible in comparison 
with the evolutionary model uncertainties \citep[which are typically $1$-$2$ 
orders of magnitude; e.g.,][]{Dominik+2012}.  

The only exception to our above conclusion are the variations of double
compact merger rates due to the uncertainty in the slope of the IMF for 
stars more massive than $1\msun$. These variations can be as high as factor 
of $\sim 3, 4, 6$ up and down for NS-NS, BH-NS and BH-BH mergers, respectively 
from our new reference model that employs the best estimates on the initial 
conditions for massive binaries.  
This is due to the change of the overall mass of the stellar mass content 
which is used to normalize the rates and to the diminishing number of double
compact object progenitors with the increasing steepness of the IMF. 

The small impact of the new distributions of binary properties on the merger
rates is somewhat counter-intuitive. This result stems from the fact that 
major changes in the initial binary parameter distributions cancel out in 
the regime important for close double compact object formation. Although, 
new distributions show a much stronger bias toward close binary 
orbits in comparison to the old distributions, the periastron distance 
distribution is similar to the one generated by old distributions in 
the regime $1.5<\log(d_{\rm per}/\rsun)<4$ (see Fig.~\ref{initdist_all}). 
This is because of the strong preference for very eccentric systems adopted 
in the older distributions, for which there is no supported in the new 
observations (mostly circular orbits).
This regime of periastron separations plays a crucial role in the formation 
of stellar-mass double compact object mergers (see Fig.~\ref{initdist_progenitors}). The 
periastron distance is key in determining when tides become important and 
when the primary star fills its Roche lobe for the first time.  Since the 
distribution of component masses is very similar, it is the periastron 
separation that largely determines the future evolution of the binary system. 
It happens (by pure coincidence) that the old prescriptions and the new 
measurements result in rather similar orbital separations at the time of the 
first mass transfer for progenitors 
of double compact objects. Additionally, the observational uncertainties of the 
new estimates of initial binary parameters for massive stars are not large 
enough to cause significant NS-NS, BH-NS and BH-BH merger rate changes. Similarly, we find that the distributions of final properties (delay-times and final component masses) are insensitive to the variations of initial conditions.  The summary of our merger rate analysis is given in Fig.~\ref{fig:uncertainties}. 
 
Our study allows to eliminate one layer of uncertainties involved in
population synthesis predictions for double compact object mergers. Such
predictions are infamous for their number of poorly constrained model 
parameters. Recent years of intensive studies of double compact object 
formation have proven that it is extremely hard (yet, not impossible) to 
identify a well posed non-degenerate problem with population synthesis. 
This fact does not reflect the intrinsic weakness of the population
synthesis method, but rather the lack of understanding of some basic
processes involved in the evolution of massive single (e.g., convection, 
rotation, supernova/core collapse) and binary stars (e.g., tidal 
interactions, common envelope evolution). 

The reduction of the modeling 
uncertainties, like the one presented here, is crucial for advancement of 
our understanding of the remaining poorly constrained physical processes 
involved in double compact object formation. The further reduction of the 
uncertainties may potentially allow gravitational-wave observatories to 
assess and constrain physical processes that are thus far beyond the 
reach of electromagnetic observations and theoretical studies.

\acknowledgements
\vspace*{0.5cm}
We would like to thank Hugues Sana, Ilya Mandel, Cole Miller, Matteo Cantiello and Colin 
Norman for many stimulating discussion that lead to this paper. We would also like to thank the anonymous referee, Ylva G{\"o}tberg, Abel Schootemeijer and Manos Zapartas for their feedback on the manuscript. 

The authors acknowledge  Carnegie Observatories for providing the
computational resources, the Caltech LIGO center for support of an
extended visit by KB  and the hospitality of University of Washington during
2014 INT workshop. 
The authors acknowledge support by the European Council (SdM) through a Marie Sklodowska-Curie 
Reintegration Fellowship (SdM), H2020-MSCA-IF-2014, project id 661502, by NASA through an
Einstein Fellowship grant (SdM), PF3-140105, the Polish Science Foundation Master 2013 
Subsidy (KB), Polish NCN grant Sonata Bis 2, DEC-2012/07/E/ST9/01360 (KB), the National Science Foundation Grant No. PHYS-1066293 (KB), the hospitality of the Aspen Center for Physics (KB) and a grant from the Simons Foundation (KB).

\bibliography{my_bib}

\end{document}